\documentclass[twocolumn,prl,superscriptaddress,noshowpacs,epsf]{revtex4}
\usepackage{graphicx}
\usepackage{epsf}
\usepackage{amsmath}
\usepackage[dvips]{color}
\usepackage{pifont}
\input{colordvi.sty}

\begin{document}
\definecolor{blue}{rgb}{0.15,0.1,0.7}
\definecolor{red}{rgb}{0.7,0.1,0.15}
\definecolor{green}{rgb}{0.15,0.7,0.15}
\newcommand{\cmt}[1]{~~[{\textcolor{blue}{\it\small #1}}]~~}
\newcommand{\red}[1]{~~[{\textcolor{red}{\it\small #1}}]~~}
\newcommand{\chk}{\cmt{$\leftarrow$ Please check}}
\newcommand{\add}{\cmt{$\leftarrow$ Please add detail}}
\newcommand{\chg}{\cmt{$\rightarrow$ How about this way?}}
\newcommand{\shorten}[1]{{\textcolor{green}{(Please consider shortening:) #1 }}}
\newcommand{\expt}[1]{{\textcolor{red}{#1}}}

\renewcommand{\onlinecite}{\cite}
\title{Leakage-induced decoherence during single electron spin manipulation in 
a double quantum dot}
\date{\today}
\author{Shi-Hua Ouyang}
\affiliation{Department of Physics and Surface Physics Laboratory
(National Key Laboratory), Fudan University, Shanghai 200433,
China}
\affiliation{Department of Applied Physics, Hong Kong Polytechnic
University, Hung Hom, Hong Kong, China}
\author{Chi-Hang Lam}
\affiliation{Department of Applied Physics, Hong Kong Polytechnic
University, Hung Hom, Hong Kong, China}
\author{J. Q. You}
\affiliation{Department of Physics and Surface Physics Laboratory
(National Key Laboratory), Fudan University, Shanghai 200433,
China}

\begin{abstract}
Coherent single electron spin oscillation in a double quantum dot
system driven by a magnetic electron spin resonance field is
studied theoretically using a Bloch-type rate equation approach. The
oscillation frequency and relaxation time obtained using typical
model parameters are consistent with experiment findings. The
dominant decoherence mechanism is identified to be a leakage current
through a Coulomb blockade barrier at a quantum dot during the
spin manipulation. Nuclear field fluctuations which induce a long
relaxation time are found to contribute only negligibly to the
decoherence despite an earlier suggestion.
\end{abstract}

 \maketitle


Because of a long decoherence time, the use of electron spins in
solid-state systems, such as semiconductor nanostructures, is
a promising approach for realizing a qubit, which is the
basic unit in quantum computing and quantum information processing
\onlinecite{Phystoday,Fiederling99,Wolf01}.
The manipulation and readout of a single electron spin
are all essential.
Solid-state implementations are expected to be most scalable and are
under intensive research.  Electrons realizing solid-state spin-based qubits are in
general confined to semiconductor quantum dots where their
numbers can be precisely controlled
\cite{Elzerman04}. The weak coupling of electron spins to the rest
of the solid allows long-lived spin states with relaxation time as
long as a few milliseconds \cite{Hanson03,Fujisawa02}. Nevertheless, this
weak coupling also makes the direct measurement of a single electron
spin challenging.
Basic operations for building up universal spin-based quantum logics
have only been realized recently.  For example, the coherent
manipulation of two electron spins in a double quantum dot (DQD)
system has been achieved experimentally \cite{Petta05}.
More importantly, the coherent rotation of {\it a single electron
spin} in a DQD system using electron spin resonance (ESR) was demonstrated
very recently by Koppens {\it et al.} \cite{Koppens Nature}.
Clear evidence of coherent oscillation of the spin state was
observed. There was also a gradual decay of the oscillation amplitude with 
a time constant of the order of 1~$\mu$s, which was
attributed to the fluctuations of nuclear fields due to the host
materials. This is consistent with the conventional belief that
nuclear fields may lead to fast relaxation of electron spins
\cite{Coish, Sham}. However, recent experiments have shown that
nuclear field fluctuations can only induce the relevant spin
transitions at a much longer time scale of $100$~$\mu$s or more at
large external magnetic fields \cite{JohnsonNature} and hence cannot
be responsible for the experimentally observed spin relaxation.

In this Letter, we derive a detailed theory for the ESR induced single
spin oscillation and the associated spin-dependent
quantum transport in a DQD system
in order to explain the experiments in \cite{Koppens Nature}. By
deriving and numerically integrating a set of Bloch-type rate
equations for the reduced density matrix elements for the DQD, we
successfully obtain the coherent oscillations of a single electron
spin when driven by an ESR field in close agreement with
experiments.
We further show that nuclear field fluctuations cannot account for
the experimentally observed decay of the spin oscillations in
contrast to the suggestion by Koppens et al.~\onlinecite{Koppens
Nature}. It is shown to be caused by a leakage current through a
Coulomb barrier during spin manipulation.  Approaches for enhancing the
coherence will be discussed.


{\it Model}~{\bf ---}~The system consists of DQD connected to two
electron reservoirs via tunneling barriers (see Fig.~1).
%
We first define the Hamiltonian for only the DQD under coherent
manipulation 
%
\begin{eqnarray}
\label{H0} H_{0}\! &=&\!\sum_{i,\sigma }E_{i}a_{i\sigma
}^{\dag}a_{i\sigma }+V_0\sum_{\sigma }(a_{L\sigma }^{\dag}a_{R\sigma
}+\rm{H.c.})
\notag\\
&+&\sum_{i}U_{i}n_{i\uparrow }n_{i\downarrow}
+U_{LR}\sum_{\sigma,\sigma'}n_{L\sigma}n_{R\sigma'} + H_{\rm mag},
\end{eqnarray}
where $i=L$ or $R$ denotes the left or right quantum dot, while
$a_{i\sigma}^{\dag}$, $a_{i\sigma}$, and $n_{i\sigma}$ are the creation,
annihilation, and number operators, respectively, for electrons at dot
$i$ with spin $\sigma$. The first four terms on the r.h.s. represent
the electron orbital energy, interdot tunneling, and intra- and
interdot Coulomb interactions, respectively. The last term describes
the interactions with magnetic fields:
\[
\label{Hmag}
H_{\rm{mag}}=\sum_{i}g\mu_B\big[\mathbf{B}_{{\rm N}i}\cdot{\mathbf{S}}_{i}
+B_{\rm ext}^{{z}}S_{{i}}^{{z}}
+B_{{i}}^{{x}}\cos(\omega_c{t})S_{{i}}^{{x}}\big],
\]
where $\mathbf{S}_{{i}}$ is the spin operator at dot $i$, while $g$
and $\mu_B$ are the electron $g$-factor and the Bohr magneton,
respectively. The nuclear magnetic field at dot $i$ due to the host
materials is denoted by $\mathbf{B}_{{\rm N}i}$. It is known to
fluctuate at a time scale of the order of 1~s, which is much longer
than that of the relevant electron transport process. They are
therefore taken as {\it stationary random} fields \cite{Nazarov}. An
external field $B_{{\rm ext}}^z$ is applied in the perpendicular
direction to generate a Zeeman splitting $
g\mu_BB_{{\rm ext}}^z$. Most interestingly, a spin at dot $i$ is
manipulated by the applied oscillating magnetic ESR field
$B_{{i}}^{{x}}\cos(\omega_c{t})$ when in resonance with the Zeeman
splitting.

The full Hamiltonian for the complete device of DQD connected to
external leads is
\begin{math}
\label{H} H\!=H_{\rm{DQD}}+\!H_{\rm{leads}}+H_{\rm{T}}.
\end{math}
%
%
The Hamiltonian of the leads is defined as
$H_{\rm{leads}}=\sum_{\alpha k\sigma}E_{\alpha k\sigma}c_{\alpha k\sigma}^+%
c_{\alpha k\sigma}$,
%
%
where $c_{\alpha k\sigma}^+$ ($c_{\alpha k\sigma}$) is the creation
(annihilation) operator of an electron with momentum $k$ and spin
$\sigma$ in lead $\alpha$ $(\alpha=l,r)$. Finally, the tunneling
coupling between the DQD and the leads is given by $H_{\rm
T}=\sum_{k\sigma}(\Omega_{l}a_{L\sigma}^{+}c_{lk\sigma}+\Omega_r
a_{R\sigma}^+c_{rk\sigma}+\rm{H.c.})$.

At least one electron is always kept in the right dot in the
experiments~\cite{Koppens Nature} by applying appropriate gate voltages. 
A second electron is transported from lead $l$ to $r$ via the dots. 
The relevant electronic states for the DQD span a seven-dimensional Hilbert
space. The basis set consists of the single-electron states
$|\!\uparrow_{R}\rangle$, $|\!\downarrow_{R}\rangle$, and five
double-electron states, namely, the double-dot triplets $|1\rangle
\equiv |T_{+}\rangle \!=\!|\!\uparrow _{L}\uparrow _{R}\rangle$,
$|2\rangle \equiv |T_{-}\rangle \!=\!|\!\downarrow _{L}\downarrow
_{R}\rangle$,
and
$|3\rangle \equiv |T_{0}\rangle\!=\!\frac{1}{\sqrt{2}}(|\!\uparrow _{L}
\downarrow_{R}\rangle +|\!\downarrow_{L}\uparrow _{R}\rangle)$,
the double-dot singlet
$|4\rangle \equiv |S_D\rangle\! =\!\frac{1}{\sqrt{2}}(|\!\uparrow _{L}
\downarrow_{R}\rangle -|\!\downarrow _{L}\uparrow _{R}\rangle)$,
and the single-dot singlet
$|5\rangle \equiv |S_S\rangle\! =\!\frac{1}{\sqrt{2}}(|\!\uparrow _{R}
\downarrow_{R}\rangle -|\!\downarrow_{R}\uparrow _{R}\rangle)$.
Single-dot triplet states are excluded due to their much higher
orbital energy \cite{Ashoori93, JohnsonNature}. In this
representation,
$H_{\rm{DQD}}$ is rewritten as
 \cite{Nazarov}
\begin{eqnarray}
\label{HDQD-triplet} &&H_{\rm{DQD}}\!=\!
\sum_{\sigma}E_{\sigma_R}|\sigma_R\rangle\langle\sigma_R|+\sum_{\beta=1}^5
E_{\beta}|\beta\rangle\langle \beta|
\notag\\
&&+\frac{g\mu_B}{\sqrt{2}}\big[(B_s^x+iB_s^y)|3\rangle\langle1|+(B_s^x-iB_s^y)|3\rangle\langle2|+\rm{H.c.}\big]
\notag\\
&&+\frac{g\mu_B}{\sqrt{2}}\big[(-B_d^x-iB_d^y)|4\rangle\langle1|+(B_d^x-iB_d^y)|4\rangle\langle2|+\rm{H.c.}\big]\notag\\
&&+V_0(|4\rangle\langle5|+|5\rangle\langle4|)+g\mu_BB_d^{z}\big(|3\rangle\langle4|+|4\rangle\langle3|\big)
\notag\\
&&+\Omega _{1}\cos (\omega _{c}t)\big[|3\rangle \langle1|+|3\rangle
\langle2|+\rm{H.c.}\big]
\notag \\
&&+\Omega _{2}\cos (\omega _{c}t)\big[-|4\rangle\langle1|+|4\rangle
\langle2|+\rm{H.c.}\big],
\end{eqnarray}
%
where
\begin{math}
\mathbf{B}_d\!=\!\frac{1}{2}(\mathbf{B}_{{\rm N}R}-\mathbf{B}_{{\rm N}L}),
~\mathbf{B}_s\!=\!\frac{1}{2}(\mathbf{B}_{{\rm N}L}+\mathbf{B}_{{\rm N}R})+
{B}^z_{{\rm ext}}\hat{\mathbf{z}},
\end{math}
and $
\Omega_{1,2}=\frac{1}{2\sqrt{2}}g\mu_B(B_{L}^{x}\pm{B}_{R}^{x}).$
We have also introduced energy levels given by
\begin{math}
E_{\sigma_R}=E_R\mp
\frac{1}{2}g\mu_B\sigma_R(B_{{\rm N}R}^z+B_{\rm ext}^z),~
E_{1,2}\!=\!E_3\mp
g\mu_BB_s^z,~E_{3,4}\!=\!E_{L}+E_R+U_{LR},~E_{5}\!=\!2E_R+U_R.
\end{math}
A critical step in the transport is the hopping to or tunneling
through the right dot. There can be a non-zero Coulomb energy
barrier
%
%
\begin{eqnarray}
\label{Delta} \Delta = E_5 - E_4 =U_R-U_{LR} - (E_L-E_R),
\end{eqnarray}
for the second electron at the right dot if the intradot repulsion $U_R$ dominates.

\begin{figure}[tbp]
\includegraphics[width=2.2in,
bbllx=10,bblly=141,bburx=592,bbury=706]{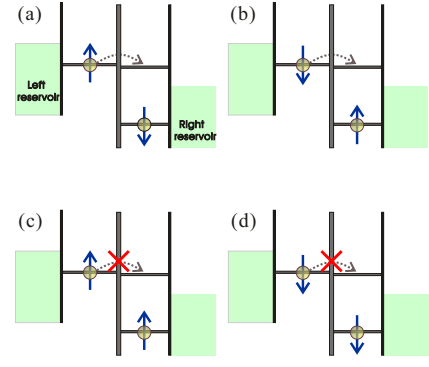} \caption{(Color
online)~Schematic diagram showing electron transfer in a DQD
structure starting at double-dot states (a)
$|\!\uparrow_L\downarrow_R\rangle$, (b)
$|\!\downarrow_L\uparrow_R\rangle$, (c)
$|\!\uparrow_L\uparrow_R\rangle$, and (d)
$|\!\downarrow_L\downarrow_R\rangle$. Electrons with $S_z=0$ in (a)
and (b) admit transport under the effects of both nuclear fields and
interdot tunneling. In contrast, transport for states with $S_z =\pm
1$ in (c) and (d) is forbidden due to the high Pauli exclusion
energy cost of the corresponding single-dot states.}
\end{figure}
%
%
%

%
{\it Coherent manipulation}~{\bf ---}~
To apply and detect the coherent rotation of an individual electron
spin, the experiment cycles repeatedly through a manipulation stage and 
a combined readout and initialization stage 
(both stages last for 1 $\mu$s)~\onlinecite{Koppens Nature}. 
In the combined readout and
initialization stage, the right dot potential is pulsed low so that
the right-dot barrier vanishes ($\Delta\simeq 0$). Also, the ESR
field is turned off ($\Omega_1=\Omega_2=B_i^{x}=0$). This
initializes the system at a spin blockade regime as will now be
explained. When a electron enters the left dot from the lead
carrying a random spin, the DQD system takes one of the four
double-electron double-dot states $|1\rangle$, $|2\rangle$,
$|3\rangle$, or $|4\rangle$ with equal probability. If the system
takes $|4\rangle$, the electron readily carries out a
spin-independent hop to the right dot as $\Delta \simeq 0$ and turns
the electrons into the single-dot state $|5\rangle$. This transition
is accounted for by the $V_0$ term in $H_{\rm{DQD}}$ in Eq.~(\ref{HDQD-triplet}). 
The electron then exits to the right lead and
this completes the transport of one electron. Alternatively,
starting at state $|3\rangle$, the $z$ components of the random
nuclear fields induce a random relative phase between the spins.
The system thus evolves into the near-degenerate state $|4\rangle$
as described by the $B^z_d$ term in $H_{\rm{DQD}}$. The electron can
then be similarly transported. As a result, the second electron can
always tunnel through the right dot for initial configurations with
$S_z=0$ [Figs.~1(a)-(b)]. In contrast, the other possible initial
states $|1\rangle$ and $|2\rangle$ are split off in energy due to
the application of a large external field $B_{\rm{ext}}^z\gg\sqrt{ <
B_{\rm N}^2 >}$ and thus are not coupled to $|3\rangle$ or $|4\rangle$.
They are hence spin blocked states and stop further transport
[Figs.~1(c)-(d)]. Therefore, starting from an arbitrary state,
current flows in general until the DQD arrives stochastically at
$|1\rangle$ or $|2\rangle$  and this completes the initialization
stage.

During the whole manipulation stage which follows, the right dot
potential is pulsed up to provide a Coulomb blockade with $\Delta \gg 0$.
A burst of oscillating ESR field is applied for a period $\tau$ just
before the end of the manipulation stage (i.e., $\Omega_1, \Omega_2 >
0$). A spin which is at resonance can be coherently rotated. The
spin blocked states $|1\rangle$ or $|2\rangle$ hence coherently
evolves into the non-spin-blocked state $|3\rangle$ or $|4\rangle$
via the $\Omega_1$ or $\Omega_2$ term in $H_{\rm{DQD}}$ in Eq.
(\ref{HDQD-triplet}). However, due to the Coulomb barrier $\Delta$
at the right dot, all transport is suggested to be completely
suppressed \cite{Koppens Nature}. We will explain later that there
is a non-negligible chance that electrons can tunnel through the
Coulomb blockade constituting an unmeasured leakage current. Other
electrons from the left lead then enter and fill the DQD again. In
all cases, there is a probability
$P(\tau)=\rho_{33}(\tau)+\rho_{44}(\tau)$ that the DQD ends up at a
non-spin-blocked state at the end of the manipulation stage. Here,
$\rho_{mn}(\tau)$ denotes a reduced density matrix element for the
DQD after manipulation.

At the subsequent read-out and initialization stage, without the
Coulomb barrier ($\Delta\simeq 0$), an electron can be transported if
it is left at a non-spin-blocked state after manipulation. This occurs
with probability $P(\tau)$. Once an electron is transported, a random
number of electrons may follow until a spin-blocked state is reached
again as explained above. It is easy to show that on average one
electron follows. As a result, the current detected in the readout
stage is
\begin{equation}
\label{I}
I_d(\tau)\!=\!\frac{2e}{T} [\rho_{33}(\tau)+\rho_{44}(\tau) ],
\end{equation}
where $e$ is the electronic charge and $T=2$~$\mu$s. In particular,
at large $\tau$ when coherence cannot be maintained,
$\rho_{33}(\tau)=\rho_{44}(\tau)=1/4$ and the current reduces to
$I_d(\infty)=e/T$.


{\it Bloch-type rate equation approach}~{\bf ---}~We now derive a
set of Bloch-type rate equations for the reduced density matrix
elements ${\rho}_{mn}$ of the DQD under the coherent manipulation by
an ESR pulse. We adopt Gurvitz {\it et al.}'s approach
\cite{Gurvitz96} in which the many-body Schr\"{o}dinger equation of
the system is reduced to quantum rate equations by integrating out
continuum reservoir states at the large voltage bias limit. We
consider the high Zeeman splitting limit in the DQD
so that spin flips caused by hyperfine interactions can be neglected.
The rate equations for the diagonal density matrix elements are
obtained after some algebra as \cite{Ouyang}
\begin{eqnarray}
\label{Deq-triplet}
&&\dot{\rho}_{00}\!=\!-4\Gamma_l\rho_{00}+\Gamma_r\rho_{55},
\notag \\
&&\dot{\rho}_{mm}\!=\!\Gamma_l\rho_{00}-i\langle m|[H_{\rm{DQD}},~\hat{\rho}]|m\rangle,
\notag \\
&&\dot{\rho}_{55}\!=\!-\Gamma_r\rho_{55}-i\langle 5|[H_{\rm{DQD}},~\hat{\rho}]|5\rangle,
\end{eqnarray}
for $m=1,2,3$ or $4$ where
$\rho_{00}=\rho_{\uparrow_R\uparrow_R}+\rho_{\downarrow_R\downarrow_R}$.
Here, $\Gamma _\alpha=2\pi \rho _\alpha\Omega _{\alpha}^{2}$ is the
transition rate for electron tunneling, while $\rho_{\alpha}$ and
$\Omega_{\alpha}$ denote, respectively, the density of states and
transition amplitude at lead $\alpha$.
The rate equations for the off-diagonal elements are
\begin{eqnarray}
\label{Oeq-triplet}
\dot{\rho}_{mn}\!&=&\!-\frac{\xi_{mn}}{2}\Gamma_r\rho_{mn}
\notag\\
&&+(1-\xi_{mn})\Gamma_l\rho_{00}-i\langle
m|[H_{\rm{DQD}},~\hat{\rho}]|n\rangle,
\end{eqnarray}
for $m, n = 1,2,3,4$ or 5 ($m \neq n$). The coefficient $\xi_{mn}$
equals one when $m$ or $n$ is $5$ and equals zero otherwise.
Any time dependence of the coefficients in Eq. (\ref{Oeq-triplet}) has been
eliminated by applying a rotating wave approximation,
which is well justified under the electron spin resonant condition
considered here \cite{RWA}.
%

%
We take typical model parameters appropriate for Koppen {\it et al.}'s
experiment. Specifically, the tunneling coupling constant between
the dots is $V_0=0.25$~$\mu$eV. The transition rates at the leads
are $\Gamma_l = \Gamma_r =10V_0$ \cite{Wiel03}. The perpendicular
field ${B}^z_{\rm ext}$ is chosen as $100$~mT. In the experiments, spins at
only one dot is at resonance with the ESR field. We assume that this
occurs at the left dot at a frequency
$\omega_c\!=\!g\mu_BB_{\rm{ext}}^z$ with a field amplitude $B_L^x$
varying from $0.74$~mT to $1.5$~mT. The field at the right dot is
non-resonant and is neglected ($B_R^x=0$). The magnitude of fluctuations of the 
nuclear fields was experimentally found to be about $2.2$~mT~\cite{KoppensScience}. 
We hence put $B_{\rm{N}R}^z=2.2$~mT and $B_{\rm{N}L}^z=0~$mT but other values of
comparable magnitude give similar results.
%

\begin{figure}[tbp]
\includegraphics[width=2.8in, bbllx=42,bblly=98,bburx=563,bbury=741]{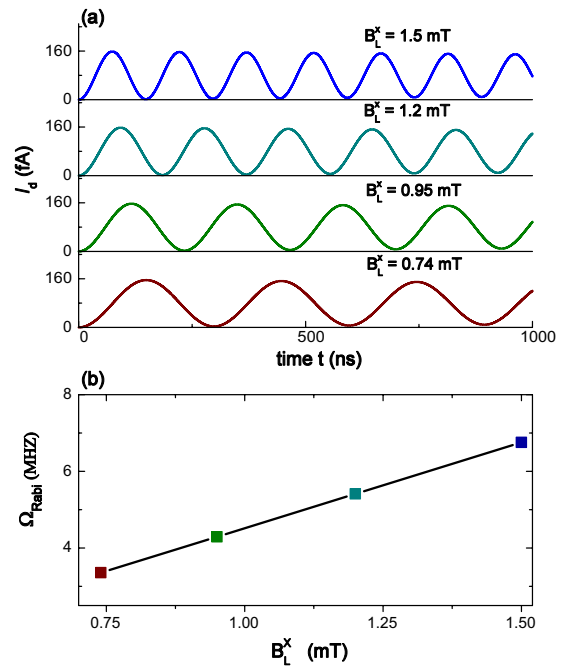}
\caption{(Color online)~(a) Plot of current $I_d(\tau)$ against ESR
burst period $\tau$ for a high Coulomb barrier $\Delta=100V_0$
showing coherent oscillations; (b)~Plot of measured frequency of
oscillation $\Omega_{\rm Rabi}$ against ESR field amplitude $B_L^x$. The
solid line represents $\Omega_{\rm Rabi}=g\mu_BB_L^x/2$.}
\end{figure}

We first consider the case of complete Coulomb blockade in the
manipulation stage as assumed in Ref.~\onlinecite{Koppens Nature} by
taking a large right-dot barrier $\Delta\!=\!100V_0$. Starting from
a spin-blocked initial state, say
$|1\rangle$, we simulate the quantum transport and manipulation
dynamics by numerically integrating the rate equations
(\ref{Deq-triplet}) and (\ref{Oeq-triplet}).  The current $I_d(\tau)$
flowing through the DQD as a function of the burst period $\tau$ is
then calculated using Eq.~(\ref{I}), and plotted in Fig.~2(a). It is
observed that $I_d(\tau)$ oscillates periodically w.r.t. $\tau$. It
can be well approximated by the analytical result at complete
Coulomb blockade given by ~\cite{Gurvitz97}:
%
$I_d(\tau)= \frac{e}{T}[1-\cos(\Omega_{\rm Rabi}\tau)]$,
%
where $\Omega_{\rm Rabi}=g\mu_BB_L^x/2$. This is due to a Rabi
oscillation of the spin state of the electron at resonance with the
ESR field. It hence implies an oscillation between the spin-blocked
and unblocked states when projected onto the singlet-triplet basis
and hence an oscillation of $I_d(\tau)$ according to Eq.~(\ref{I}).
The oscillation frequency $\Omega_{\rm Rabi}$ measured from our numerical
data agrees with the exact relation above [Fig.~2(b)] and in
particular has a linear dependence on the amplitude of the ESR field
$B_L^x$ in agreement with experiments. In this case, we observe practically 
no decay of the current oscillations as the Rabi oscillations
have almost constant amplitudes. This is however in sharp contrast to
the gradual decay from experiments. Therefore, our result shows
clearly that decay cannot be explained by the quasi-static
nuclear fields in the presence of complete Coulomb blockade as have
been suggested by Koppens {\it et al.}~\cite{Koppens Nature}.
This is in fact consistent with a recent experiment
\cite{JohnsonNature} showing that the relevant transition between
triplet-singlet spin states in DQD as induced by nuclear fields
takes much longer durations of about $100$~$\mu$s and 1~ms at
external fields of $30$~mT and $150$~mT, respectively. Furthermore,
other spin-relaxation mechanisms including spin-orbit interactions
and hyperfine interactions have even longer time scales
\cite{kroutvar04,Golovach} and similarly cannot account for the
decay of the current oscillations.



Our theory agrees much better with experiments when a tunneling
leakage current through the Coulomb blockade barrier is considered.
Here, we assume a more realistic value for the Coulomb barrier
$\Delta\!=\!20V_0 = 5$~$\mu$eV which will be further explained below.
The corresponding simulation result is shown in Fig.~3. We find that
the Rabi oscillations damp gradually during the ESR burst at a rate
in good agreement with experiments. A smaller $\Delta$ would lead to
an even faster decay. The oscillation period $\Omega_{\rm Rabi}$
depends linearly on the amplitude of the ESR field as before.  The
current $I_d(\tau)$ oscillates around the limiting value
$I_d(\infty)=e/T=80$~fA in agreement with experiments. Finally, the
Coulomb barrier $\Delta$ for an experimental device is limited by
the fact that the right dot potential cannot be pulsed arbitrarily
high because one electron has to remain all the time.
From Eq.~(\ref{Delta}), one can see that $\Delta$ does not only
depend on the difference between the intra- and interdot Coulomb
repulsions, but also on the difference of the orbital levels for the
two dots. It can be shown that the value of $\Delta$ used here is
then a reasonable one for the experiment when these two differences
are comparable.

%
\begin{figure}[tbp]
\includegraphics[width=2.8in,
bbllx=33,bblly=151,bburx=567,bbury=694]{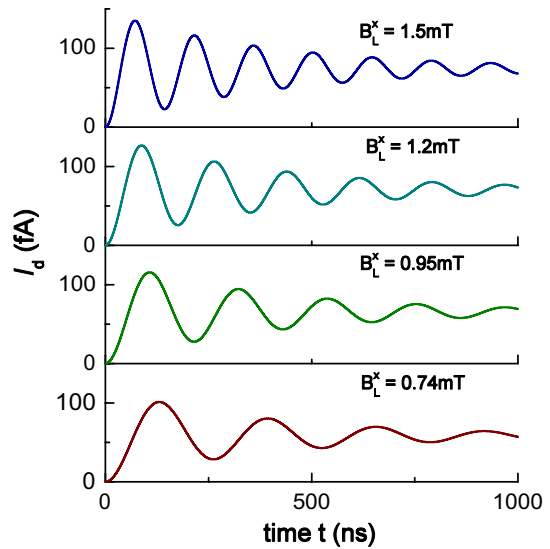} \caption{(Color
online)~Same as in Fig.~2(a), but for a small Coulomb barrier
$\Delta=20V_0$ showing coherent oscillation with damping.}
\end{figure}
%



In conclusion, we have investigated the coherent rotation of a
single electron spin in a DQD system driven by an ESR field. A set
of Bloch-type rate equations have been derived analytically and
integrated numerically. We then obtain a current oscillation as a
direct consequence of a coherent Rabi oscillation of an individual
electron spin state in agreement with experiments. We also verify
that the frequency of the oscillation has a linear dependence on
the amplitude of the ESR field. A leakage current through a Coulomb
blockade barrier during the manipulation stage is identified as the
dominant mechanism of decoherence. Nuclear field fluctuations which
induce a relaxation time much longer than the manipulation period are
shown to contribute negligibly to the decoherence despite an
earlier suggestion. 
Our detailed quantitative theory
which allows the identification of the relevant decoherence
mechanism is of great importance. Based on our result, it is
evident that the coherence can be improved by suppressing the leakage
current. This can be achieved by enhancing the Coulomb barrier
$\Delta$ by fabricating a smaller right dot in a similar device, or
by decreasing the orbital level difference via tuning the
appropriate gate voltages. We hope that our results can motivate
further experimental investigations.


\begin{acknowledgments}
This work was supported by PCSIRT, SRFDP, NFRPC grant No.~2006CB921205,
and the National Natural Science Foundation of China 
grant No.~10625416.
\end{acknowledgments}


\end{document}